# Sparsity based sub-wavelength imaging with partially incoherent light via quadratic compressed sensing

Yoav Shechtman[1], Yonina C. Eldar[2], Alexander Szameit[1] and Mordechai Segev[1]

[1]*Physics Department and Solid State Institute, Technion, Haifa 3200, Israel*
[2]*Department of Electrical Engineering, Technion, Haifa 3200, Israel*

**Abstract:** We demonstrate that sub-wavelength optical images borne on partially-spatially-incoherent light can be recovered, from their far-field or from the blurred image, given the prior knowledge that the image is sparse, and only that. The reconstruction method relies on the recently demonstrated sparsity-based sub-wavelength imaging. However, for partially-spatially-incoherent light, the relation between the measurements and the image is quadratic, yielding non-convex measurement equations that do not conform to previously used techniques. Consequently, we demonstrate new algorithmic methodology, referred to as quadratic compressed sensing, which can be applied to a range of other problems involving information recovery from partial correlation measurements, including when the correlation function has local dependencies. Specifically for microscopy, this method can be readily extended to white light microscopes with the additional knowledge of the light source spectrum.

**OCIS codes:** (180.0180) Microscopy; (110.4980) Partial coherence in imaging; (100.3010) Image reconstruction techniques

**Introduction**

The resolution of an optical imaging system has traditionally been bounded by Abbe's limit – setting the smallest resolvable feature in an imaged object to be about the size of $\lambda/2$, with $\lambda$ being the optical wavelength of the light illuminating the object [1]. Indeed, when imaging an object containing sub-wavelength features, these features become blurred and unresolvable. The reason for this blurring is that only waves containing spatial frequencies lower than $1/\lambda$ propagate through the medium, while higher spatial frequencies become evanescent and decay exponentially in a homogeneous medium. Such waves are greatly attenuated over a propagation distance larger than several $\lambda$, hence the nature of Maxwell's equation in homogeneous media acts practically as an ideal low-pass filter [1]. Over the years, there have been many attempts to recover information with a resolution exceeding the diffraction limit. Some of these techniques have become standard commercial tools, e.g. the scanning near-field optical microscope [2], and methods based on light-emitting molecules [3]. Other approaches offer great promise but their technology still requires major improvements, e.g., devices based on negative-index materials [4,5,6], or scanning with nanoholes [7] or with a sub-wavelength hot-spot [8]. However, one common feature limits almost all of these "hardware solutions" for sub-wavelength imaging: they are not single-exposure experiments, hence they cannot work in real time, because they either require scanning [2,7,8], or multiple experiments [3]. Needless to say, it would be highly desirable to have sub-wavelength imaging technology operating at ultrafast scales, which is the time scale of many important chemical reactions.

During the past decades, there have been attempts to recover sub-wavelength information through bandwidth extrapolation algorithms. These techniques utilize the fact that the 2D Fourier transform of a spatially bounded function (the electromagnetic field) is an analytic function, hence measuring the far-field at very high precision could, in principle, allow recovery of the information contained in the evanescent waves, which has never reached the far field. However, these attempts were not successful, as they are extremely sensitive to noise in the measured data and to the assumptions made on the prior knowledge about the sub-wavelength information (see discussion in [1, page 165]).

Recently, we have demonstrated a method to overcome the diffraction limit [9], by using prior knowledge that the imaged object is sparse in a known basis (say, in the near-field). The methodology is based on bandwidth extrapolation, where the extrapolating function is chosen such that it yields the sparsest solution (image) that is consistent with the measurements. Mathematically, the problem was posed in matrix representation, and solved using efficient tools borrowed from the realm of signal processing, specifically – methods for finding sparse solutions to underdetermined systems of equations [10,11]. Experimentally, using coherent laser light, we were recently able to recover sub-wavelength features as small as 100nm using 532nm wavelength [12]. However, it is highly desirable to develop such techniques also for the incoherent case – simply because most microscopes work with incoherent light. Consequently, we have demonstrated sparsity-based super-resolution with completely spatially-incoherent quasi-monochromatic light [13]. However, taking the technique of [13] into the sub-wavelength regime (as we did for coherent light [12]) has limited applicability, because it uses the Optical Transfer Function (OTF) which assumes that the correlation distance is much smaller than any feature in the optical object. The idea would work when light emitting molecules are distributed on the object itself, but would not work in microscopes using incoherent light. This is because microscopes use lenses (or mirrors) to collect the incoherent illumination; hence the correlation distance at the object plane cannot be smaller than the diffraction limit of the collecting lens. Consequently, when using incoherent light to illuminate sub-wavelength optical information, the optical features can be smaller, comparable, or larger than the correlation distance of the light in the object plane, implying that the OTF cannot represent the propagation of the light adequately. In such cases, which are most common in microscopy, the light incident upon sub-wavelength optical objects must be treated as partially-spatially-incoherent.

Here, we demonstrate theoretically the reconstruction of objects containing sub-wavelength features, by using the prior knowledge that the objects are sparse in real space, i.e. containing only a few non-zero pixels. Unlike the two extreme cases of coherence that we previously dealt with [9,13], the partially-incoherent case cannot be represented by a transfer function, which translates into linear measurements in the electric-field [9] or in the intensity [13], for fully coherent and for completely incoherent light, respectively. Instead, data resulting from the partially-incoherent case is represented by a more complicated relation, depending on the mutual intensity [14]. As such, the signal processing theory and methods, which are well established for sparse solutions to a set of linear equations, cannot be directly applied here. We therefore develop the underlying theory for the reconstruction of sub-wavelength images borne on partially-spatially-incoherent light, and show that the resulting problem calls for finding a sparse solution to a set of quadratic equations. We consequently propose a heuristic method to find a sparse solution to this set of quadratic equations, a problem which we name Quadratic Compressed Sensing, and demonstrate (via simulations) the reconstruction of sparse sub-wavelength objects illuminated by partially incoherent light using this approach.

Our method can be extended to work under broad-band illumination ("white light"), provided that the spectral composition of the light is known. Hence, the techniques demonstrated here can be generalized to work with white light microscopes, which are the most commonly used microscopes. Finally, this concept of reconstructing information from a small subset of measurements of its correlation function is universal, having a broad variety of applications in optics and beyond. Examples range from Fourier transform spectroscopy (FTIR), Frequency Resolved Optical Gating (FROG) and other methods relying on interference, to equivalent examples with any other kind of signals. Perhaps most intriguingly, the concept established here could pave the way to the recovery of quantum states from a small subset of measurements of their correlation function, given only that the original quantum information is sparse in a known basis.

**Propagation of images borne on partially-spatially-coherent light**

An optical imaging system is typically designed to image the intensity structure of an EM field (henceforth the "object") placed at some input plane, to some other place ("output plane") where the output intensity structure ("image") is measured with a camera (or displayed on a screen). Here, we are interested in the case where the object contains sub-wavelength features. In the extreme case where the transverse correlation distance (sometimes called "transverse coherence length") of the light is much smaller than the smallest feature of the object, the light can be assumed to be completely spatially incoherent, and therefore the imaging process can be described by the simple OTF for the intensity [1]. In that case, one can recover the information carried by the spatial frequencies that are smaller than the cut-off frequency by simply dividing the Fourier transform of the image by the OTF. This is a standard deconvolution procedure, which works perfectly well when the OTF is nonzero. Similarly, for the fully-coherent case, i.e. when the transverse coherence length of the light is much larger than the largest features in the object, the imaging process is described by the amplitude transfer function (ATF) of the system, sometimes also called the Coherent Transfer Function [1]. One can then recover the field amplitude of the object – up to the cut-off frequency - by dividing the Fourier transform of the image-amplitude by the ATF. In both the coherent and incoherent cases, all information about the frequencies higher than the cut-off is lost.

Recently, we suggested and demonstrated a method to recover the information carried by high frequencies that were cut-off by the imaging process – for both the coherent and incoherent cases [9,12,13], by exploiting the object's sparsity in the near field.

Here, we suggest a way to recover sub-wavelength information carried by light that is neither completely coherent nor completely incoherent – but with some partial spatial coherence. We utilize prior knowledge that the information (the object intensity) is sparse in real space, and only that. As we shall show, the case of partially-incoherent illumination leads to an interesting mathematical problem that is fundamentally different than the extreme coherence cases – and requires a novel method for recovering the information.

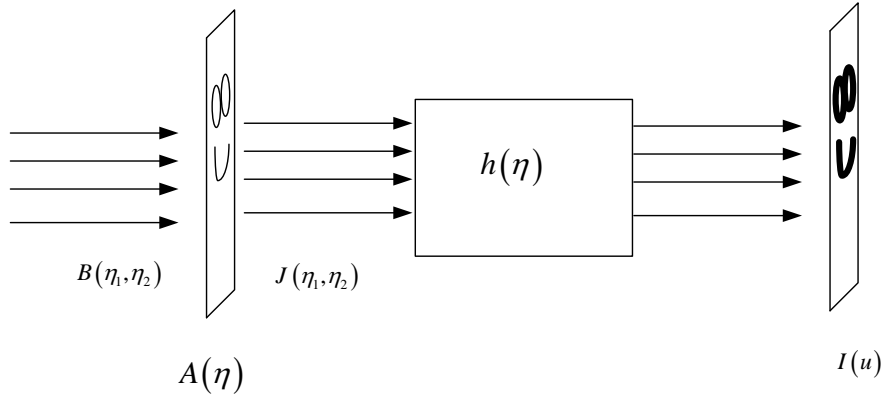

**Fig. 1**: Setting for imaging using partially spatially coherent light.

Consider the setting sketched in Fig. 1. A thin optical object with complex transmittance function $A$, placed at plane $z=0$, is illuminated by quasi-monochromatic partially-spatially-coherent light. For simplicity, we analyze here one dimensional objects, hence $A$ is a function of $\eta$, the transverse coordinate at the input plane; however, the extension to two transverse dimensions is straightforward. The object is imaged using an optical system with a coherent impulse response $h(u;\eta)$, $u$ being the transverse coordinate at the image plane. Note that $h$ is a function of the coordinates of both the input and the output planes. In other words, an object consisting of a single narrow peak $A(\eta) = \delta(\eta - \eta_0)$ would yield the image $A_{image}(u) = h(u;\eta_0)$, had it been illuminated by perfectly spatially-coherent illumination. We measure the light intensity distribution $I(u) = \langle |A_{image}(u,t)|^2 \rangle$ at the image plane, where $\langle \cdot \rangle$ denotes the time-average, and the averaging is carried out over the response time of the camera (or the eye), which is assumed to be much longer than the characteristic fluctuation time of the incoherent light. From the measurements of $I(u)$, we wish to retrieve the original intensity distribution of the object, and, if possible, also its phase.

For most physical systems, the transmittance function $A(\eta)$ defining the object does not vary with time. The light incident upon the object is quasi-monochromatic and partially-spatially-incoherent, that is, its complex electric field $u(\eta,t)|_{z=0^-}$ is fluctuating randomly with some characteristic time that is much longer than one cycle of the optical frequency. The light at the input plane ($z=0$) is neither fully spatially coherent nor completely spatially incoherent.

Hence, the field amplitudes at two different points $\eta_1$ and $\eta_2$ in plane $z=0$ exhibit some correlation, described by the mutual intensity function $B(\eta_1,\eta_2) \triangleq \langle u(\eta_1,t)u^*(\eta_2,t) \rangle \big|_{z=0^-}$. Note that the ability to recover the phase, even when the impulse response is hypothetically a delta function, is limited due to the incomplete spatial coherence of the light. Namely, the partially incoherent light is characterized by a transverse correlation distance $L_c$ between any two points in the transverse plane; for two points separated by $\eta < L_c$ the relative phase is well defined, whereas any two points separated by $\eta > L_c$ are not phase-correlated. Hence, phase information that varies over distances much shorter than $L_c$ is lost, and cannot be recovered.

As explained above, the light travels through the optical system, until reaching the image plane at which the intensity measurements are conducted. The instantaneous electric field of the light travels as a coherent field, and is therefore described by the coherent impulse response of the system, $h(u;\eta)$. When the impulse response of the optical system is space-invariant, with proper coordinate scaling we can write $h(u;\eta) \triangleq h(u-\eta)$, so that the intensity of light reaching the image plane can be written as (see page 313 in [14]):

$$I(u) = \iint h(u-\eta_1)h^*(u-\eta_2)J(\eta_1,\eta_2)d\eta_1 d\eta_2 \qquad (1)$$

where $J(\eta_1,\eta_2)$ is the mutual intensity function immediately after the optical object ($z=0^+$). Since the optical object $A(\eta)$ does not vary in time, one can write:

$$\begin{aligned}J(\eta_1,\eta_2) &= \langle A(\eta_1,t)u(\eta_1,t)A^*(\eta_2,t)u^*(\eta_2,t) \rangle = \\ &= A(\eta_1)A^*(\eta_2)\langle u(\eta_1,t)u^*(\eta_2,t) \rangle = A(\eta_1)A^*(\eta_2)B(\eta_1,\eta_2)\end{aligned} \qquad (2)$$

For most sources emitting partially-spatially-incoherent light (e.g., the sun, thermal sources, a laser beam passing through a rotating diffuser), $B(\eta_1,\eta_2)$ is a function of the coordinate difference $|\eta_1-\eta_2|$ only. Under the above circumstances, Eq. (1) can be rewritten as:

$$I(u) = \iint h(u-\eta_1)h^*(u-\eta_2)A(\eta_1)A^*(\eta_2)B(|\eta_1-\eta_2|)d\eta_1 d\eta_2. \qquad (3)$$

An example illustrating the effect of the mutual intensity function is depicted in Fig. 2, showing a 1D object comprising of two peaks spaced by $0.6\lambda$ (Fig. 2a,b) and by $\lambda$ (Fig. 2c,d), respectively. The impulse response function $h(u)$ of the optical system is taken as $h(x) = \dfrac{h_0 \sin(2\pi x/\lambda)}{x}$, corresponding to an ideal low-pass-filter with a spatial cutoff frequency of $1/\lambda$. The peaks have either the same phase (Fig. 2a,c) or opposite phases (Fig. 2b,d). Figure 2 shows the intensity at the image plane, $I(u)$, for three cases: fully coherent (dotted line), completely incoherent light (dashed line), and partially-incoherent light (dash-dotted) with the mutual intensity function being a Gaussian with $0.55\lambda$ Full Width Half Maximum (FWHM), i.e., transverse correlation distance is $L_c = 0.55\lambda$.

Three key points should be emphasized in Fig. 2. First, the image is a smeared (blurred) version of the object, regardless of the coherence of the light, demonstrating the low-pass nature of the system. Second, the effect of the coherence of the light on the image is much more pronounced when the peaks are close together, specifically having sub-wavelength spacing. Third, when the optical information varies on a sub-wavelength scale (Fig. 2a,b), the partially-incoherent image (dash-dotted) is considerably different than both the coherent and incoherent images. This implies that using either the fully coherent or the fully incoherent approximations to recover the object from its measured image would not yield good recovering. Hence, in physical settings where the optical information varies on a sub-wavelength scale, it is essential to employ a method dealing specifically with partially-spatially-coherent light.

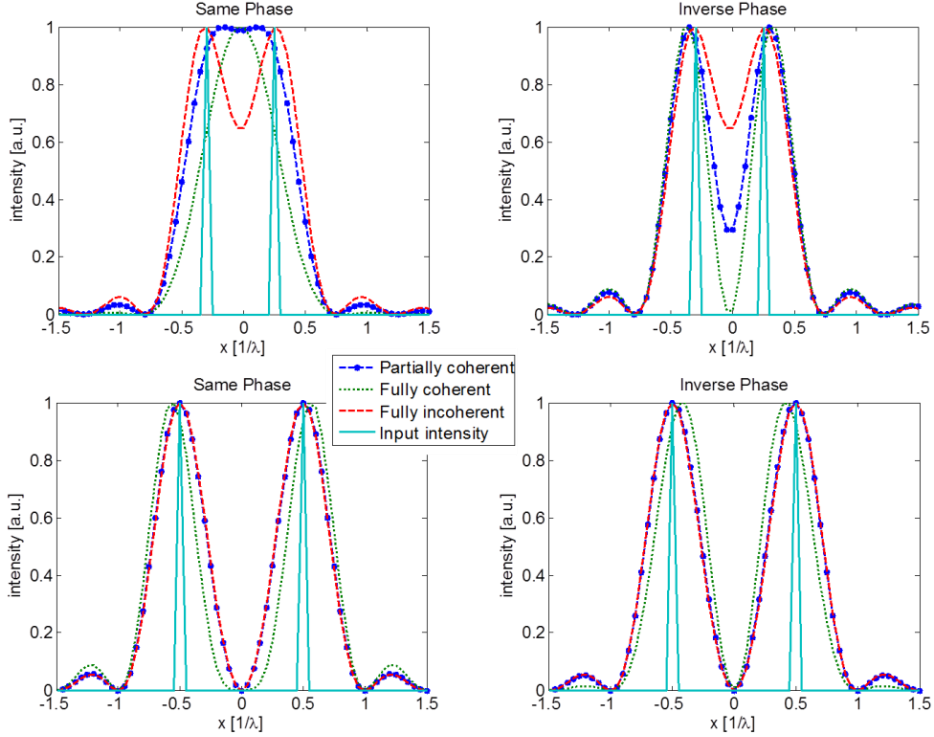

**Fig. 2**: Illustration of the effect of mutual-intensity width on output intensity, for 2 equal phase (a) or inverse phase (b) spikes, separated by 0.55λ (same/inverse phase). In (c) and (d) the spikes are 1λ apart. Coherent light corresponds to a mutual-intensity width of infinity; incoherent light corresponds to mutual-intensity width of 0. The partially coherent mutual-intensity width here is 0.55λ FWHM. The resolution is better for equal phase spikes when using incoherent light (a), and for inverse phase spikes – using coherent light (b). Note that the effect of the coherence of the light is much stronger for closely (sub- λ) separated peaks, and that the partially-incoherent image in that case (dash-dotted in (a) and (b)) is considerably different than both the coherent and incoherent images.

**Matrix representation**

Our goal is to solve Eq. (3). Namely, to find the transmittance function $A(\eta)$ given $h(\eta)$, $B(|\eta_1 - \eta_2|)$ and the measured intensity $I(u)$. In order to use our sparsity-based reconstruction algorithm, we first discretize the problem and rewrite Eq. (3) in matrix form. This yields

$$I(u) = \sum_i \sum_j h_u(\eta_i) h_u^*(\eta_j) B(\eta_i - \eta_j) A(\eta_i) A^*(\eta_j) \qquad (4)$$

using $h_u(\eta) \triangleq h(u-\eta) \cdot d\eta$, where $d\eta$, the discretization parameter, is selected to be smaller than the features in the image, with the purpose of making the discrete problem similar enough to the true analog problem from which it originated. Eq. (4) can be written in a compact form as:

$$y_u = a^* M_u a \qquad (5)$$

where we define $M_{u,ij} \triangleq h_u(\eta_i) h_u^*(\eta_j) B(\eta_i - \eta_j)$. The vector $y_u$ represents the measured intensity, where the index $u$ refers to the location of the $u^{th}$ measurement in the image plane. We denote the set of all measurements by $U$.

Eq. (5) describes a quadratic relation between the object $a$ and the $u^{th}$ measurement, related by a matrix $M_u$, for all $u \in U$. The size of $y_u$ is determined by the number of measurements, and the size of $M_u$ is determined by the discretization step.

Our goal is now to find a vector $a$ that includes the optical information at the input plane, or to "invert" Eq. (5).

**Lack of uniqueness and Sparsity prior**

Unfortunately, under even small amounts of noise, there is no unique vector $a$ that solves Eq. (5). In other words, there are many different vectors $a$ that would yield the same measurements $y_u$ (within the noise level), after being 'smeared' by the optical system. Our goal is to find the true one. One can notice immediately that uniqueness can never be guaranteed. For example, if $\tilde{a}$ is a solution to Eq. (5), then so is $-\tilde{a}$, or in fact any $\tilde{a} \cdot e^{i\varphi}$ for any phase $\varphi$. We therefore limit ourselves to reconstruction of $\tilde{a}$ up to a global phase constant. owever, even when allowing a global phase ambiguity, there is still no uniqueness, under a small amount of noise. The reason for this is the low-pass nature of the system: It can be shown (based on [1, chapter 7.2.3]), that the spectrum of the output image intensity is bounded in k-space by $2\nu_c$, where $\nu_c$ is the cutoff frequency of the coherent transfer function of the system, irrespective of the spectrum of the input image or of the coherence properties of the impinging light. This causes inherent loss of information, of the same sort that was dealt with previously in the two extreme cases of completely coherent and completely incoherent imaging [9,13].

The low-pass nature of the system can also be inferred by examining the form of the $M_u$ matrices, as will be shown below. Note that the cut-off frequency of the image implies a theoretical upper bound for the sampling rate (or a lower bound for the distance between two samples), given by the sampling theorem, beyond which no new frequency information will be obtained. This is referred to as the Nyquist rate and it is in our case equal to $4\nu_c$, which in an ideal diffraction-limited optical system corresponds to the spacing between samples of λ/4. We shall see later that, indeed, increasing the sampling rate above this rate does not improve the results significantly.

Therefore, in our current problem, in order to overcome the lack of uniqueness - in the spirit of the fully coherent and fully incoherent extreme cases - we assume a sparse model of the input field (in real space), meaning that the input field comprises of only a few nonzero values in space. Sparsity-based algorithms are known to handle linear problems very well, that is, problems of the form $y = Ma$ where $y$ is the measurement vector, $a$ is the information we wish to recover, and $M$ is the transformation relating the two. Here, our problem is not linear, as manifested by the matrix representation of Eq. 5. Hence, we cannot directly harness the theoretical knowledge about linear sparse problems to solve the current problem. Nevertheless, we can still use the prior knowledge about the sparsity of $a(\eta)$ to aid in the reconstruction process.

**Optimization problem: formulation and solution method**

We are looking for the sparsest input field $a(\eta)$ that is consistent with the measurements $y(u)$. In addition, in a passive optical system with no gain, the overall output power cannot be larger than the overall input power. This translates mathematically to solving the following optimization problem:

$$(P) \min_{a} \|a\|_0 \text{ subject to } |a^* M_u a - y_u|^2 \leq \varepsilon \quad u \in U \qquad (6)$$
$$a^* a \geq y^* y$$

where $\|a\|_0$ denotes the 0 'norm' of $a$, or simply the number of nonzero elements in $a$. $U$ is the set of measurements taken at the image plane. The consistency parameter $\varepsilon$ is determined by the typical amount of noise added by the system.

This is not a convex problem, due to both the objective function (which stands for the sparsity requirements) and to the quadratic consistency and energy constraints. In general, when seeking a sparse solution that is consistent with a set of measurements, the sparsity requirement is never convex. When the measurement vector is related linearly to the signal vector (so-called "linear measurements" in signal processing), many methods have been developed to approximate the sparsest solution (for a review of these methods, see [11]). However, in our case the relation between the measurements vector and the signal vector is quadratic (Eq. (5)). To the best of our knowledge, there are no prior results addressing such a problem, hence we have to find a new method to approximate the sparsest solution under these circumstances of "quadratic measurements".

We solve the quadratic problem posed by Eq. 6 as follows. We define a matrix $X \triangleq aa^*$, such that the non-convex consistency constraint in (6) may be re-written as a convex constraint on the matrix $X$ as follows: $a^* M_u a \triangleq Trace(M_u X)$. Note that we must also add the requirement that $X$ can be written as $X \triangleq aa^*$ for some vector $a$ – which is really what we are after. This requirement means that $X$ is positive-semi-definite (PSD), and a Rank 1 matrix. The Rank 1 requirement is not convex, so instead of adding it as a constraint, we attempt to minimize the rank, subject to the remaining constraints, as shown below in (7).

Next, we deal with the sparsity constraint - sparsity of $a$ implies row-sparsity of $X$, i.e. if $a$ has $k$ nonzero elements ($k$-sparse), then $X$ has $k$ rows that contain nonzero elements, and each of these rows is also $k$-sparse. The row-sparsity of $X$ is promoted by adding a constraint on the upper bound of the mixed-1,2 norm, i.e. $\sum_a \sqrt{\sum_b X_{ab}^2} < \zeta$, with $\zeta$ being a parameter derived from the degree of assumed sparsity of the object. This requirement promotes a solution $X$ that is sparse in rows, and furthermore each row in itself is a sparse vector. This is the kind of structure that we are looking for in a solution matrix $X$ that should represent the unknown sparse vector $a$. The energy conservation constraint is reformulated as $tr(X) \geq y^* y$. We can now express our problem (P) as a minimization problem in the matrix $X$.

To summarize, we are looking for a matrix $X$ that is consistent with the measurements, PSD, row-sparse and rank 1. So the optimization problem we wish to solve is the following:

$$\operatorname*{argmin}_{X} \; Rank(X) \; s.t.$$

$$\sum_{a}\left(\sum_{b} X_{ab}^{2}\right)^{1/2} \leq \zeta$$
$$|tr(M_u X) - y_u| \leq \varepsilon \; \forall u \in U \quad (7)$$
$$tr(X) \geq y^* y$$
$$X \geq 0$$

The algorithm we develop to try to solve (7) is based on an iterative rank minimization heuristic method [15], that attempts to find a low-rank matrix consistent with a set of convex constraints, by minimizing the log-det (the logarithm of the determinant) of the matrix. In other words, the low-rank objective is replaced by the function $\log\det(X + \delta I)$ with $\delta$ being a small regularization parameter. This function is not convex, however the method suggested in [15] is shown to converge to a local minimum. The requested properties of our solution are therefore incorporated as convex constraints into the log-det algorithm.

Summarizing, the optimization problem we actually solve is the following:

$$\operatorname*{argmin}_{X} \; \log\det(X + \delta I) \; s.t.$$

$$\sum_{a}\left(\sum_{b} X_{ab}^{2}\right)^{1/2} \leq \zeta$$
$$|tr(M_u X) - y_u| \leq \varepsilon \quad (8)$$
$$tr(X) \geq tr(y \cdot y^T)$$
$$X \geq 0$$

The basic iteration of the log-det heuristic method is performed by solving the following convex optimization problem:

$$X_{k+1} = \operatorname*{argmin}_{X \in C} \; Tr(X_k + \delta I)^{-1} X \quad (9)$$

In our case, $C$ is the set of convex constraints written in (8). In addition, to further promote a sparse solution, at each iteration a thresholding step is performed, in which the smallest elements in the diagonal of $X$ are located, and the intersecting rows and columns are taken out of the possible support in the next iteration. This is added as a convex support constraint to the next iteration. Each iteration can be solved using standard convex optimization tools.

Once a low-rank $X$ is found (empirically, $X$ is approximately Rank 1 at the end of the process), $aa^*$ is determined as the best rank-1 approximation of $X$, using the Singular Value Decomposition (SVD). If the rank minimization process succeeds, i.e. the solution matrix is close to rank-1, then of course its rank-1 approximation will also be row-sparse as required. From there, the approximation of $a$ is obtained. Algorithm 1 describes the algorithm in detail.

The following parameters are used in the algorithm:
- $\varepsilon$ - Determined by the noise added by the optical system. This could, in principle, be calibrated using a known reference object.
- $\zeta$ - Determined by the assumed sparsity of the object.
- $\delta$ - A small regularization parameter – in practice the algorithm's performance is not very sensitive to it.

- $T$, $\Delta T$ - Define the thresholding strength, determined by the system's noise.

The actual values used in the simulations are given in the Algorithm-evaluation section.

---

**Goal**: Find a sparse solution $\hat{a} \in C^n$ consistent with the measurements

$$\left| a^* M_u a - y_u \right|^2 \leq \varepsilon \quad \forall u \in U$$

**Parameters**: $\delta, \zeta, \varepsilon, T, \Delta T$

---

1. **Initialization**: $X^0 = I^{n \times n}$, off-support=$\varnothing$

2. **Solve** $X^k = \underset{X}{\operatorname{argmin}} \, Tr\left( X^{k-1} + \delta I \right)^{-1} X$, under the following constraints:

$$\sum_a \left( \sum_b X_{ab}^2 \right)^{1/2} \leq \zeta$$
$$\left| tr(M_u X) - y_u \right| \leq \varepsilon$$
$$tr(X) \geq tr(y \cdot y^T)$$
$$X \geq 0$$
$$X(\text{off-support}) = 0$$

3. **Thresholding**: Define: $Diag = Diagonal(X^k)$. For all indices $i$ for which $Diag(i) < T \cdot \max |Diag|$, add column $i$ and row $i$ to the off-support in the next iteration. If the off-support was not updated, increase $T$ iteratively by $\Delta T$ until off-support is updated.

4. **Repeat** stages 2-3 until there is no solution to stage 2. This means that the off-support is too large – the result $X^k$ is taken as the output of the last successful iteration. If by this time $X^k$ is still not close enough to rank 1 (Defined by $\dfrac{s_1}{s_2} > 10^3$ where $s_j$ is the j$^{th}$ singular value of $X^k$), continue iterating with previous successful off-support.

5. **Final solution**: Calculate the Singular Value Decomposition of the resulting $X$, so that $USV^T = X$. The requested $\hat{a}$ is then given by $\hat{a} = U_1 \cdot s_1$ where $U_1$ is the column of $U$ corresponding to the largest singular value $s_1$.

**Algorithm 1**

**The transfer matrices $M_u$**

The set of transfer matrices $M_u$ describing the imaging process are determined by both the mutual intensity function of the light incident on the object plane, and the coherent impulse response of the optical system. These are defined as follows: $M_{u,ij} \triangleq h_u(\eta_i) h_u^*(\eta_j) B(\eta_i - \eta_j)$, where $h_u(\eta) \triangleq h(u-\eta) \cdot d\eta$, and $h(u-\eta)$ is the mirrored and discretized coherent impulse response of the system, shifted by u. Each matrix $M_u$ is created digitally by:

$$M_u = h_u h_u^* .* B \qquad (10)$$

where $B$ is the Toeplitz matrix representing the mutual intensity right before the object plane. For example, for a Gaussian mutual intensity distribution, each row in $B$ is a shifted Gaussian, and the symbol '.*' denotes element-wise multiplication.

As an example, we show a subset of $M_u$ matrices in Fig. 3. Each matrix $M_u$ defines the relation between the unknown object $a$ and the measurement $y_u$ (related through Eq. (5)). Observing Fig. 3, it is apparent that each measurement is a function of the different values of $a$, centered around the measurement point. There are two physical quantities that affect this dependence. The first quantity is the coherence of the incident light, which affects the effective width of each row in $M_u$. In the extreme (physically unattainable) case where the light is completely incoherent, each $M_u$ is a diagonal matrix, and then each measurement is a function only of $|a|^2$, i.e. of the intensity at each point in the object. As the coherence of the light is increased, each measurement becomes a function of terms mixing pairs of values of $a$ farther and farther apart (i.e. terms of the form $a_i a_j^*, i \neq j$). The other quantity affecting $M_u$ is the coherent impulse response function of the system which affects the effective spreading of each $M_u$ along the *diagonal* direction. The more spread-out the coherent impulse response of the system is, the more spread out each $M_u$ will be along the diagonal, therefore each measurement $y_u$ will be a mixture by values of $a$ farther away from it. Since the width of the coherent impulse response is at best bounded from below by ~$\lambda/2$ which corresponds to the diffraction limit, each measurement $y_u$ will always be a weighted sum of values in a region of $a$, which would always lead to smearing of the signal.

This 'matrix' viewpoint is indeed consistent with the physical cutoff frequency acting on the output signal: the output field of the partially-coherent light propagating in the system must be bounded in k-space, due to the low-pass nature of light propagation. An example of this fact can be seen in Fig. 4c, where the output image is bounded in k-space by $2/\lambda$.

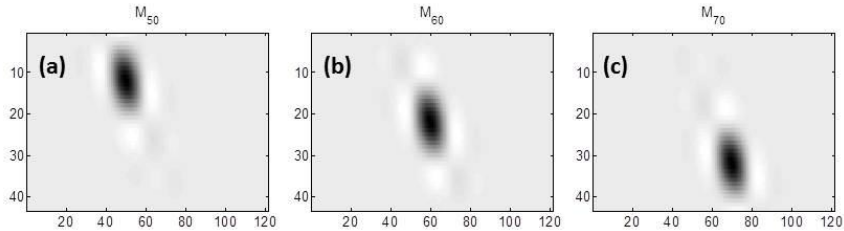

**Fig. 3**: Example subset of $M_u$ matrices (dark indicates high values), corresponding to a Gaussian mutual-intensity function, and an impulse response of the form of sinc(x). Shown are the matrices $M_{50}$ (a), $M_{60}$ (b), and $M_{70}$ (c).

**Numerical reconstruction**

We test the proposed reconstruction algorithm numerically, using sparse objects in real space, namely – a few spikes at varying distances and amplitudes. A typical reconstruction simulation is depicted in Fig. 4. In Figs. 4a and 4b an object comprising of 4 spikes distanced $0.5\lambda$ apart is illuminated by partially-incoherent light, with a Gaussian mutual intensity function that is $0.55\lambda$ wide (FWHM). The light is then propagating through a diffraction-limited imaging system, and the intensity of the blurred image is measured. The coherent impulse response of the system $h(x)$ is approximated as a sinc function, corresponding to an ideal low-pass filter, yielding: $h(x) = \dfrac{h_0 \sin(2\pi x/\lambda)}{x}$. [In reality, the impulse response of an optical imaging system, especially at high numerical apertures, is considerably more complicated, accounting for aberrations etc.; experimentally, one would actually measure the coherent impulse response of the optical system (amplitude and phase), and substitute it numerically into the algorithm]. To demonstrate the robustness of our technique, we add noise at the level of 40dB (1%) to the measured intensity distribution of the blurred image, i.e. $y_{Noisy} = y + \nu$ where $\nu$ is random Gaussian noise and $\dfrac{\|\nu\|_2}{\|y\|_2} = 10^{-2}$ (reasonable for an optical system). Using the "measured" data, we attempt reconstructing the object, using only the detected (smeared) image, and knowledge that the object is sparse in real space. Figure 4a shows the reconstructed object using the sparsity prior. For comparison, we show in Fig. 4b the reconstructed object without using the knowledge about the sparsity of the input image. Namely, the thresholding procedure is removed from the log-det heuristic iterations, and the small mixed 1-2 constraint is removed as well. The sparsity-based reconstructed image of Fig. 4a is virtually identical to the true sparse object.

The sparsity prior knowledge is essential in order to obtain successful reconstruction. This is illustrated in Fig. 4b, where sparsity is not used; hence the reconstructed object is very different than the input optical information (the object we wish to recover).

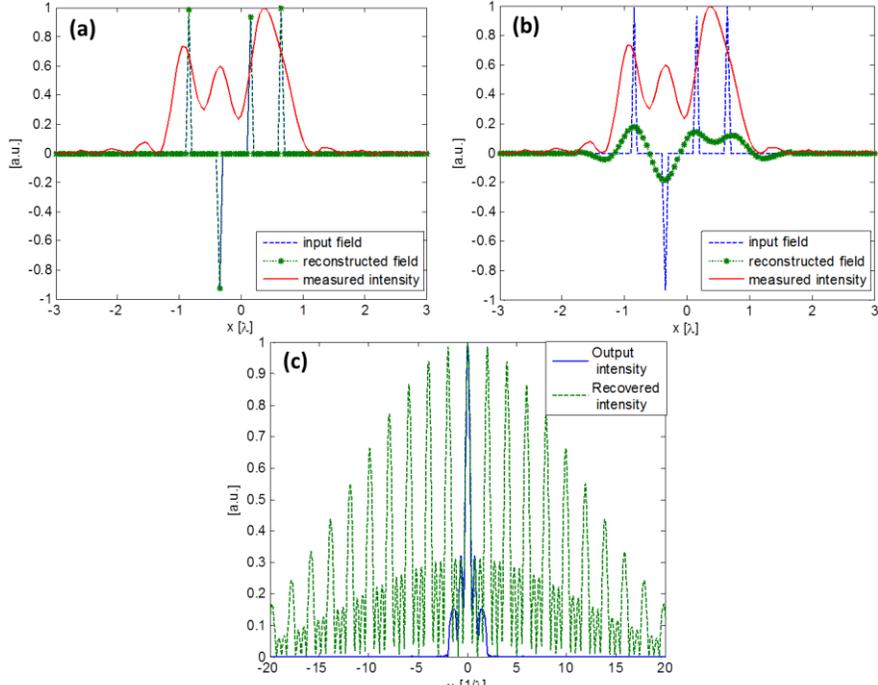

**Fig. 4**: Reconstruction of a sparse object consisting of 4 spikes at different phases (±π), distanced 0.5λ apart. The mutual intensity function of the light impinging on the object is a Gaussian with FWHM of 0.55λ. The coherent impulse response of the optical system is taken as $h(x)=\frac{1}{x}h_0\sin\left(\frac{2\pi x}{\lambda}\right)$.

(a) Sparsity-based reconstruction, which yields reconstruction that is practically identical to the true input field (dashed-blue). (b) Reconstruction under the same assumptions, without using the sparsity prior Both reconstructions (dotted green, marked with *) are consistent with the measurements (red). The sparsity-based method of (a) yields a reconstruction virtually identical to the initial object, whereas the reconstruction that does not use sparsity, shown in (b), cannot retrieve any of the fine features of the object. The corresponding bandwidth extrapolation is shown in (c) – the Fourier spectrum of the recovered intensity (dashed green) contains information much higher than the output intensity, which is limited by the physical cutoff frequency at 2/λ (in blue).

**Algorithm evaluation**

Several aspects of the algorithm's performance are tested numerically. First of all, the necessity of the sparsity assumption is verified. Fig. 5 shows the reconstruction error as a function of different noise levels, both for the sparsity-based reconstruction, using the proposed algorithm, and for a reconstruction that does not assume sparsity. The latter is obtained by using the same algorithm, but omitting the sparsity requirements (i.e. the mixed norm bound) and the thresholding iterations. The recovery error is defined as follows:

$$\varepsilon_{\text{rec}} = \frac{\|a_{\text{rec}} - a_{\text{true}}\|_2}{\|a_{\text{true}}\|_2} \tag{11}$$

where $a_{\text{rec}}$ is the recovered input signal, and $a_{\text{true}}$ is the true input signal, after they have been convolved with a Gaussian of FWHM of 0.1λ (the results are fairly robust to this parameter), in order to allow a small deviation from the true locations, and yet to penalize for larger deviations. The input signal is comprised of 3 peaks, distanced 0.35λ from each-other, each with a random phase (of ±π), and amplitude of 10±ν where ν is a normally distributed random variable, i.e. $v \sim N(0,1)$. The number of measurements taken is 121. However, due to the low-pass filter nature of the system, the measured signal is bounded in k-space by $2/\lambda$. Therefore, increasing the number of measurements above the Nyquist rate (in this example this translates to 25 samples), does not add more frequency information, although it does have the effect of improving the reconstruction in terms of noise-robustness.

Each point in the plot is the average over 50 reconstructions. In all simulations the mutual intensity function of the light incident on the object is a Gaussian with FWHM of $0.55\lambda$. Both reconstruction methods perform better as noise is reduced, but the results clearly show the advantage of exploiting the sparsity of the optical information at the input plane. Another aspect that is tested is the effect of the number of peaks on the reconstruction error. This time, the input signal is comprised of several peaks, equally spaced from one another, within a range of $1.3\lambda$. The other parameters are identical to the ones used to produce Fig. 4. Figure 6 shows the average reconstruction error as a function of the number of peaks. The results show that as the signal becomes less sparse (higher peak density), the error increases.

Finally, although the main purpose of this work is not to reduce the number of samples necessary for reconstruction, the effect of the number of samples taken in the image plane is also examined. This is shown in Fig. 7, where the reconstruction error is plotted as a function of the number of equally-spaced measurements in the image plane. The tested scenario is that of 3 peaks, separated by $0.45\lambda$, with other parameters identical to those of Fig. 5. Above a number of measurements comparable to the Nyquist rate (~25 measurements), the reconstruction does not improve dramatically. The problem of bandwidth extrapolation is not solved by adding more samples, since the physical cutoff frequency of the optical system obviously does not depend on this quantity.

In all of the simulations described above, the parameter values (See algorithm box 1) are selected as follows: $\delta = 10^{-3}$, $\zeta = 1.1 \cdot \sum_a \left( \sum_b \tilde{X}_{ab}^2 \right)^{1/2}$, $\varepsilon = \max_u |\tilde{y}_u - y_u|$, $T = 10^{-1}$, $\Delta T = 10^{-2}$ where $\tilde{X} \triangleq \tilde{a}\tilde{a}^*$ with $\tilde{a}$ being the true input signal. $y_u$ are the measurements without noise, and $\tilde{y}_u$ are the noisy measurements.

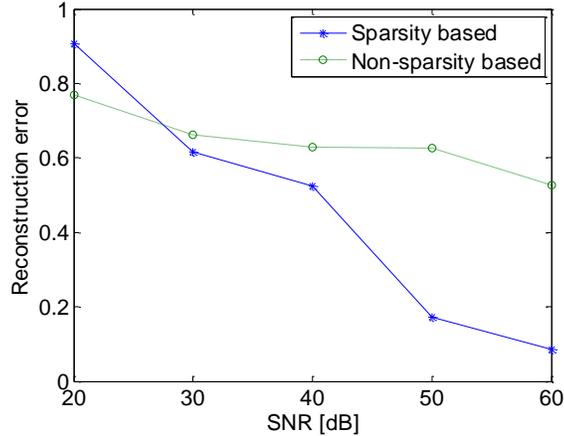

**Fig. 5:** Reconstruction error as a function of noise in the measurements, for sparsity-based reconstruction (marked with *) and for non-sparsity based reconstruction (dashed). The advantage of sparsity based reconstruction is clear, especially under low noise levels.

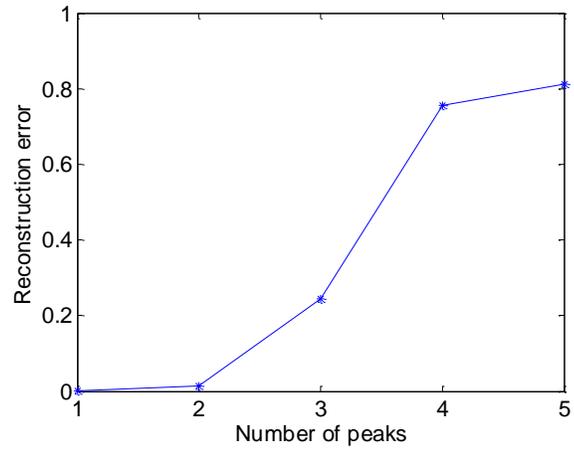

**Fig. 6:** Reconstruction error as a function of the number of peaks. The peaks are equally spaced inside a range of 1.3λ. Each point is averaged over 100 reconstructions.

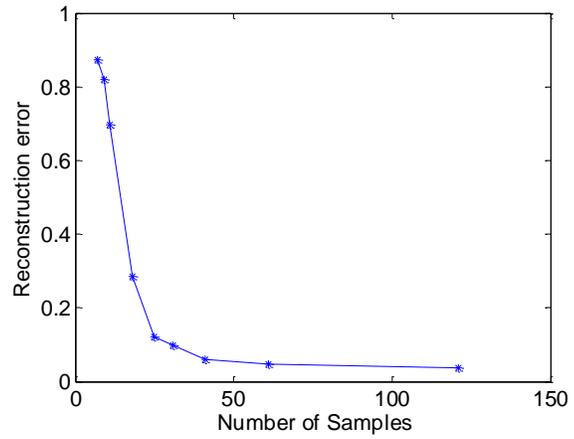

**Fig. 7:** Reconstruction error as a function of the number of samples in the image plane. The samples are equally spaced. Each point in the plot is averaged over 100 reconstructions.

**Summary**

We presented a sparsity-based technique to recover sub-wavelength features of optical objects illuminated by partially-incoherent light, by using intensity measurements of the blurred image, and exploiting the knowledge that the object is sparse (and only that). The method is based on finding a sparse solution to a system of quadratic equations. Mathematically, this is done by minimizing the rank of a positive semi-definite matrix, under the constraints arising from the physical imaging system.

In a more general scope, the techniques described here show how to recover information from partial (truncated) measurements of the correlation function. The immediate application of this general idea for sub-wavelength optical imaging with partially-spatially-incoherent light can be readily extended to include light with partial temporal coherence, with the knowledge of the spectral distribution of the light. However, the concept of sparsity-based reconstruction of data from partial correlation measurements holds the promise for other areas as well. An immediate example that comes to mind is for FTIR spectroscopy, where our group has indeed demonstrated the ability to distinguish between adjacent atomic lines using measurements of the truncated autocorrelation function. Perhaps most intriguing is the possibility of implementing these ideas on a quantum information system where the measurements are always correlation measurements. Recent work on Compressed Sensing of quantum information has demonstrated signal recovery from sub-Nyquist sampling [16,17], but has not treated the specific issues of signal recovery from low-pass filtered measurements, or truncated correlation function. Would it be possible to recover information on entangled states from partial (truncated) measurements of the correlation function? Would these ideas also work for higher-order correlation functions? Can these ideas be applied to quantum tomography, quantum photolithography, and quantum cryptography? We expect much future research on sparsity-based techniques to recover information from partial / truncated correlation measurements.